\documentstyle[12pt,epsf]{article}
\newlength{\abstractwidth}
\renewcommand{\thefootnote}{\fnsymbol{footnote}}
\renewcommand{\thanks}[1]{\footnote{#1}} 
\newcommand{\starttext}{
\setcounter{footnote}{0}
\renewcommand{\thefootnote}{\arabic{footnote}}}
\renewcommand{\theequation}{\thesection.\arabic{equation}}
\newcommand{\be}{\begin{equation}}
\newcommand{\bea}{\begin{eqnarray}}
\newcommand{\eea}{\end{eqnarray}}
\newcommand{\beq}{\begin{equation}}
\newcommand{\ee}{\end{equation}}
\newcommand{\eeq}{\end{equation}}

\newcommand{\<}{\langle}

\renewcommand{\>}{\rangle}
\def\ba{\begin{eqnarray}}
\def\ea{\end{eqnarray}}

\def\14{{1\over4}}
\def\12{{1 \over 2}}

\def\h3{h^{3\over 2}}

\def\>{\rangle}
\def\<{\langle}

\def\cc{cosmological constant}
\def\ap{Anthropic Principle}

\def\sb{supersymmetry breaking}
\def\sbs{supersymmetry breaking scale}
\def\bdg{Banks Dine Gorbatov}
\def\aa{anthropically acceptable}

\begin{document}
\renewcommand{\theequation}{\thesection.\arabic{equation}}
\begin{titlepage}
\bigskip
\rightline{SU-ITP 02-11} \rightline{hep-th/0204027}

\bigskip\bigskip\bigskip\bigskip

\centerline{\Large \bf {Supersymmetry Breaking in the Anthropic
Landscape }}

\bigskip\bigskip
\bigskip\bigskip

\centerline{\it L. Susskind  }
\medskip
\centerline{Department of Physics} \centerline{Stanford
University} \centerline{Stanford, CA 94305-4060}
\medskip
\medskip

\bigskip\bigskip
\begin{abstract}
In this paper I attempt to address a serious criticism of the
``Anthropic Landscape" and ``Discretuum" approach to cosmology,
leveled by Banks, Dine and Gorbatov. I argue that in this new and
unfamiliar setting, the gauge Hierarchy may not favor low energy
supersymmetry. In a added note some considerations of Douglas
which substantially strengthen the argument are explained.
\end{abstract}

\end{titlepage}
\starttext \baselineskip=18pt \setcounter{footnote}{0}

\setcounter{equation}{0}
\section{The Banks Dine Gorbatov Argument }
Let's begin by reviewing a successful use of the anthropic
principle. In 1987  Steven Weinberg predicted that
if the anthropic principle was correct, the cosmological constant
should not be very much smaller than the bound provided by galaxy
formation \cite{sw}. The argument is straightforward: a much
smaller value would require fine tuning of the type which the
anthropic principle was supposed to eliminate. Since the anthropic
upper bound was only a couple of orders of magnitude above the
empirical upper bound, Weinberg argued that the anthropic
principle predicted that a \cc \ would soon be discovered. And it
was.

It is worth noting that there are also anthropic bounds on the
weak scale that might be strong enough to require a gauge
hierarchy like the one we actually observe. Increasing the Higgs
expectation value keeping everything else fixed would among other
things, increase the strength of gravity, making stars, galaxies
and the universe evolve much faster. Alternatively we could try to
keep particle masses fixed by decreasing Yukawa couplings. This
would decrease the strength of weak interactions, thereby having
many effects on the creation of heavy elements in stars as well as
the primary mechanism for dispersing the elements, namely,
supernovae.

Banks, Dine and Gorbatov (BDG) \cite{bdg} have recently argued
that similar logic can be applied to proton stability and that the
exceptionally long life of the proton falsifies the \ap . The
context of the \bdg \ argument is the ``Landscape" or
``Discretuum" hypothesis \cite{bp, kklt, ls, ad}. They argue
(correctly I think) that the anthropic bound on the proton
lifetime is about $10^{17}$ years. Therefore  the vast majority of
\it anthropically acceptable \rm landscape sites have proton
lifetimes many orders of magnitude shorter than the experimental
lower limit $~10^{32}$ years.

One way out of the \bdg \ argument is to note that in the
non-supersymmetric standard model, there is no mechanism for
proton decay. If the scale signaling the breakdown of the
standard model is high enough, there is no problem. Thus BDG
begin by first arguing that the vast majority of \aa \ vacua
have a very low supersymmetry breaking scale. Their argument goes
as follows:

Without supersymmetry both the \cc \ and the Higgs mass scale,
$\mu$ must be fine tuned. The combined fine tuning is about one
part in $10^{150}$. If the \sbs \ is called $M$ then the natural
scale for radiative corrections to the \cc \ is of order $M^4$. If
the actual \cc \ is $\lambda$ then the likelihood of radiative
corrections cancelling and leaving the small value $\lambda$ is of
order
\be
P(M,\lambda) =\lambda / M^4
\label{new}
\ee
 Thus making $M$ as small as possible will
yield the largest number of \aa \ vacua. Similarly they argue that
since the Higgs mass is quadratically sensitive to $M$ the actual
measure of fine tuning is
\be
P(M,\mu,\lambda) =\lambda \mu ^2/ M^6.
\label{fintun}
\ee
 For $M $ of order the Planck mass we get $10^{-150}$ but for $M$ of order the
weak scale, the fine tuning is only of order $10^{-60}$. There is
therefore a very strong statistical bias toward low energy \sb .

But low energy supersymmetry breaking is not without its problems.
Many of the things which the standard model solved so
neatly--proton stability, neutral strangeness changing currents
and the like-- are neither automatic nor especially natural in
supersymmetric extensions.  In particular, dimension 5 baryon
violating operators are dangerous. Without some special
non-generic symmetry, the proton lifetime in supersymmetric
theories is several orders of magnitude too small. I might point
out that dimension 4 operators can also occur in generic
supersymmetry models. These lead to extremely short lifetimes
unless R-parity forbids them. But unlike the dimension 5
operators, the dimension 4 operators are so bad that they can be
ruled out anthropically.

The \bdg \ criticism is quite serious in my opinion, and needs to be
addressed. In this note I will argue that \bdg \ overlooked one
important statistical factor in their analysis which, when
included, may change the conclusion. Here is the way I think the
argument should go:

What we want to compute is the conditional probability that the
supersymmetry breaking scale is $M$, given that the \cc \ and weak
scale are in the anthropic range. This is not the function $P$ in
\ref{fintun}. Rather the function $P$ is the conditional
probability that the \cc \ and weak scale are in the allowed
range, given that the \sbs \ is $M$. The two distributions are
related by Bayes' theorem. Define the probability that the \cc \
and weak scale are in the allowed range, given that the \sbs \ is
$M$ to be
$$
P(\lambda,\mu |M)
$$
and the probability that the supersymmetry
breaking scale is $M$, given that the \cc \ and weak scale are in
the anthropic range to be
$$
P(M|\lambda,\mu).
$$
Also define the unconditional probabilities for given \sbs \ and
for the values $\lambda,\mu$ to be

$$
P(M), \ \ P(\lambda,\mu).
$$

Then Bayes' theorem relates these probabilities \be
P(M|\lambda,\mu)= P(\lambda,\mu|M) P(M)/P(\lambda,\mu).
\label{bayes} \ee

Thus in comparing the likelihood of different values of $M$ the
factor $1/M^6$ should be multiplied by $P(M)$. The factor in the
denominator of \ref{bayes} can be ignored since it is independent
of $M$. The question is, what is the value of the unconditional
probability $P(M)$?

I believe it is likely that $P(M)$ goes to zero as $M$ tends to
zero. The reason is that the conditions for supersymmetry
generally pick out a subspace of the landscape whose
dimensionality may be a good deal lower than the number of
available dimensions including those that parameterize
supersymmetry breaking.  A reasonable guess is that $P(M)$ should
go to zero as a power of $M$ in which case it could overwhelm the
factor $1/M^6$. To illustrate the point I will concoct an
illustrative example within a
 framework similar to that of \cite{kklt}. The \sb
\ mechanism and the discrete fine tunings of the \cc \ and Higgs
scale take place in different sectors of the theory. In particular
the \sb \ sector is located at the infrared end of a warped
compactification while the tunings of the \cc \ and $\mu$ are done
through the choice of fluxes in the ultraviolet part of the
compactification manifold. The only modification of \cite{kklt} is to suppose
that there are $n$ throats instead of just one.

In \cite{kklt} the mechanism for supersymmetry breaking is one or
more antibranes placed at the end of the throat. The \sbs \ is the
combined mass of the branes. Thus
\be
M = N M_D
\ee
where $M_D$ is the antibrane mass and N is the number of
antibranes.

If there are $n$ throats a given total number of antibranes can be
distributed among the throats in many ways. For example $n_1$
antibranes can be placed in the first throat, $n_2$ in the second
throat and so on. The number of ways of partitioning the mass $M$
among $n$ throats is of order $M^{n-1}$. Thus if the number of
throats exceeds six, the distribution favors high energy \sb .
 \cite{Denef}.

Which kind of throat structure--many or one-- is favored by
statistics of the landscape? This may be a delicate question.
With no condition of supersymmetry breaking, one or even no
throats may dominate. But you can't live in a supersymmetric
world. Given a scale of supersymmetry breaking, the statistics may
be tilted toward many throats just because there are more ways to
break supersymmetry if there are many throats.

If supersymmetry breaking is at a high scale then it would seem
that the apparent unification of coupling constants is accidental.
Dimopoulos and Arkani-Hamed have discussed a possible framework in
which supersymmetry can be broken at a high scale but the coupling
constant unification not be destroyed \cite{ahd}.

Finally, I should point out that the problem of the QCD theta
parameter, pointed out by \bdg \ is not obviously eliminated by
the considerations of this paper but the problem may be less
serious if supersymmetry is broken at a very high scale. String
theory contains many axion-like objects but they generally belong
to super-multiplets that contain scalar moduli. If supersymmetry
is an approximate symmetry then fixing the moduli will result in a
large axion mass. But if supersymmetry is broken at a high scale
and the moduli are fixed at a lower scale then there is no reason
why the axions must get a large mass.

If nothing else, I hope this note shows that the measure on the
landscape is a very subtle issue that demands better quantitative
methods of the kind being developed by Douglas and collaborators
\cite{ad}. The whole idea of fine-tuning has to be re-evaluated
and redefined in this new unfamiliar framework.

After this paper was written I became aware of the fact that
Denef, Douglas and Florea have discussed the problem of
supersymmetry breaking in the Landscape. These authors find that
high scale supersymmetry breaking may be favored \cite{Denef}.

 \setcounter{equation}{0}
\section{Note Added}
After this paper was written M Douglas put out a paper addressing
the same problem \cite{newd}. Douglas' conclusion is a good deal stronger than
mine. The difference lies in formulas \ref{new} and \ref{fintun}.
Douglas would replace \ref{new} by
\be
P(M,\lambda) =\lambda / M_p^4
\label{newer}
\ee
where $M_p$ is the  Planck or string scale. The argument assumes
that the cosmological constant satisfies a formula like
\be
\lambda = \lambda_0 + RC \ee where the first term is classical
supergravity contribution and $RC$ is the exact radiative
correction. Assume that the discretuum gives a spectrum of values
of $\lambda_0$ that is more or less uniform over the range from
$-M_p^4 $ to  $+M_p^4 $ (My mistake was to assume the spectrum is
bounded by $\pm M^4$ instead of $\pm M_p^4$). Also assume that the
radiative corrections are of order $M^4$. If no more than that is
known then we can write
\be
\lambda = x M_p^4 + y M^4
\ee
where $x$ and $y$ have a dense spectrum lying between $1$ and
$-1.$ Douglas gives strong arguments why the spectrum of $x$ is
featureless with no singularity at $x=0.$ Under these
circumstances \ref{newer} is correct. Thus Douglas concludes that
\ref{fintun} should be replaced by
\be
P(M,\mu,\lambda) ={\lambda \mu ^2 \over M^2 M_p^4}.
\label{newfintun}
\ee
In other words the only the gauge hierarchy fine-tuning favors low
energy supersymmetry breaking. In this case we would need fewer
powers of $M$ in $P(M)$ to overwhelm the bias toward low scale
breaking.

However in conversations with Douglas and Dimopoulos we realized
that the same logic may apply to the hierarchy as to the
cosmological constant in which case the bias would be nonexistent
altogether. In this case $P(M,\mu,\lambda)$ would favor high scale
breaking even with no enhancement from $P(M)$.

 \setcounter{equation}{0}
\section{Acknowledgements}
I would like to thank Tom Banks, Michael Dine,  Ben Freivogel,
Savas Dimopoulos Steve Shenker, Scott Thomas Shamit Kachru and
Michael Douglas for very useful discussions.


\begin{thebibliography}{999}

\bibitem{bdg} Tom Banks, Michael Dine, Elie Gorbatov,
Is There A String Theory Landscape, hep-th/0309170

\bibitem{sw} Steven Weinberg, ANTHROPIC BOUND ON THE COSMOLOGICAL CONSTANT,
Phys.Rev.Lett.59:2607,1987


\bibitem{bp}  Raphael Bousso, Joseph Polchinski,
Quantization of Four-form Fluxes and Dynamical Neutralization of
the Cosmological Constant, hep-th/0004134, JHEP 0006 (2000) 006

\bibitem{kklt} Shamit Kachru, Renata Kallosh, Andrei Linde, Sandip P.
Trivedi, de Sitter Vacua in String Theory, hep-th/0301240


\bibitem{ls}  Leonard Susskind, The Anthropic Landscape of String Theory,
hep-th/0302219


\bibitem{ad}  Michael R. Douglas, The statistics of string/M vacua,
 hep-th/0303194,  JHEP 0305 (2003) 046


Sujay Ashok, Michael R. Douglas, Counting Flux Vacua,
hep-th/0307049

 Michael R. Douglas, Bernard Shiffman, Steve Zelditch,
 Critical points and supersymmetric vacua
math.CV/0402326

\bibitem{ahd} Nima Arkani-Hamed, Savas Dimopoulos,
 Supersymmetric Unification Without Low Energy Supersymmetry
 And Signatures for Fine-Tuning at the LHC,
hep-th/0405159


\bibitem{Denef}
F.~Denef, M.~R.~Douglas and B.~Florea, Building a better
racetrack, arXiv:hep-th/0404257

\bibitem{newd} Michael R. Douglas, Statistical analysis of the
supersymmetry breaking scale,
hep-th/0405279

\end{thebibliography}
\end{document}